\documentclass[aps,pre,twocolumn,showpacs,superscriptaddress,floatfix,%
              showkeys,reprint]%
    {revtex4-1}
\usepackage{graphicx}
\usepackage{bm}
\usepackage{amsfonts,amssymb,amsmath}
\usepackage{color}
\usepackage[colorlinks=true,%
            citecolor=black,%
            linkcolor=black,%
            urlcolor=blue]{hyperref}

\usepackage{graphicx}
\usepackage{amsmath,amsfonts}
\usepackage[english]{babel}
\usepackage{bm}    %% For bold math symbols
\usepackage{eepic} %% For defining plot symbols \fancyplus etc

\newcommand{\be}{\begin{equation}}
\newcommand{\ee}{\end{equation}}
\newcommand{\<}{\langle}
\renewcommand{\>}{\rangle}

\def\spose#1{\hbox to 0pt{#1\hss}}
\def\ltapprox{\mathrel{\spose{\lower 3pt\hbox{$\mathchar"218$}}
 \raise 2.0pt\hbox{$\mathchar"13C$}}}
\def\gtapprox{\mathrel{\spose{\lower 3pt\hbox{$\mathchar"218$}}
 \raise 2.0pt\hbox{$\mathchar"13E$}}}

\newcommand{\scrm}{{\cal M}}

\newcommand{\scrq}{{\cal Q}}

\newsavebox{\fancyplusb}

\savebox{\fancyplusb}(2,2){
\thinlines
\put(0,-1){\line(0,1){2}}
\put(-1,-0.5){\line(0,1){1}}
\put(-1,0){\line(1,0){2}}
\put(1,-0.5){\line(0,1){1}}
\put(-0.5,-1){\line(1,0){1}}
\put(-0.5,1){\line(1,0){1}}
}

\newtheorem{conjecture}{Conjecture}

\newcommand{\arxiv}[1]{\href{http://arxiv.org/abs/#1}{\texttt{arXiv:#1}}}

\begin{document}

\title{Duality and the universality class of the
     three-state Potts antiferromagnet \\ on plane quadrangulations} 

%% http://tex.stackexchange.com/questions/5805/revtex-4-1-multiple-affiliations
\author{Jian--Ping Lv}
\email{phys.lv@gmail.com}
\affiliation{Department of Physics, Anhui Normal University,
             Wuhu 241000, China}
\affiliation{Anhui Province Key Laboratory of Optoelectric Materials Science 
             and Technology, Wuhu 241000, China}
\author{Youjin Deng}
\email{yjdeng@ustc.edu.cn}
\affiliation{Hefei National Laboratory for Physical Sciences at Microscale
   and Department of Modern Physics,
   University of Science and Technology of China,
   Hefei, Anhui 230026, China}
\affiliation{CAS Center for Excellence and Synergetic Innovation Center
   in Quantum Information and Quantum Physics, University of Science and 
   Technology of China, Hefei, Anhui 230026, China} 
\author{Jesper Lykke Jacobsen}
\email{jesper.jacobsen@ens.fr}
\affiliation{Laboratoire de Physique Th\'eorique,
 D\'epartement de Physique de l'ENS, \'Ecole Normale Sup\'erieure,
 Sorbonne Universit\'e, CNRS, PSL Research University, 75005 Paris, France}
\affiliation{Sorbonne Universit\'e, \'Ecole Normale
 Sup\'erieure, CNRS, Laboratoire de Physique Th\'eorique (LPT ENS),
 75005 Paris, France}
%% https://www.ipht.fr/en/Phocea-SPhT/ast_visu_spht.php?id_ast=486
\affiliation{Institut de Physique Th\'eorique, CEA Saclay,
 91191 Gif-sur-Yvette, France}
\author{Jes\'us Salas}
\email{jsalas@math.uc3m.es}
 \affiliation{Grupo de Modelizaci\'on, Simulaci\'on Num\'erica y Matem\'atica 
   Industrial, Universidad Carlos III de Madrid, Avenida de la Universidad 30,
   28911 Legan\'es, Spain}
\affiliation{
  Grupo de Teor\'{\i}as de Campos y F\'{\i}sica Estad\'{\i}stica, 
  Instituto Gregorio Mill\'an, UC3M, Unidad Asociada al IEM-CSIC, 
  Madrid, Spain} 
\author{Alan D. Sokal}
\email{sokal@nyu.edu}
\affiliation{Department of Physics, New York University,
      4 Washington Place, New York, NY 10003, USA}
\affiliation{Department of Mathematics, University College London,
      Gower Street, London WC1E 6BT, UK}
      
\date{December 21, 2017; Revised  February 6, 2018}

\begin{abstract}
We provide a new criterion based on graph duality to predict whether the 
3-state Potts antiferromagnet on a plane quadrangulation has a zero- or
finite-temperature critical point, and its universality class.
The former case occurs for quadrangulations of self-dual type,
and the zero-temperature critical point has central charge $c=1$.
The latter case occurs for quadrangulations of non-self-dual type,
and the critical point belongs to the universality class of the
3-state Potts ferromagnet.
We have tested this criterion against high-precision computations
on four lattices of each type, with very good agreement.
We have also found that the Wang--Swendsen--Koteck\'y algorithm
has no critical slowing-down in the former case,
and critical slowing-down in the latter.
\end{abstract}



\maketitle

\section{Introduction}

Ever since Kramers and Wannier's \cite{Kramers_41} pioneering work
on the two-dimensional (2D) Ising model,
the concept of duality has led to important insights
in statistical mechanics and quantum field theory \cite{Savit_80}
and more recently also in string theory \cite{Polchinski_96}.
The purpose of this Rapid Communication is to show an unusual 
application of duality
to the study of the 3-state Potts antiferromagnet (AF)
on a class of 2D lattices.

The $q$-state Potts model \cite{Potts_52,Wu_82+84}
plays a key role in the theory of critical phenomena,
especially in 2D \cite{Baxter_book,Nienhuis_84,DiFrancesco_97},
and has applications to various condensed-matter systems \cite{Wu_82+84}.
Ferromagnetic Potts models are by now fairly well understood,
owing to universality; but the behavior of AF Potts models
depends strongly on the microscopic lattice structure,
so that many basic questions
about the phase diagram and critical exponents  
must be investigated case-by-case.
One expects that for each lattice ${\cal L}$ there
exists a value $q_c({\cal L})$ [possibly noninteger]
such that for $q > q_c({\cal L})$  the model has an exponential decay
of correlations at all temperatures~$T$ including zero,
while for $q = q_c({\cal L})$ the model has a zero-temperature critical point.
The first task, for any lattice, is thus to determine~$q_c$.

Some 2D AF models at $T=0$  
have the remarkable property that they can be mapped exactly onto a ``height''
model (in general vector-valued) 
\cite{Henley_unpub,Kondev_96,Burton_Henley_97,Salas_98,Jacobsen_09}.
Since the height model must either be in a ``smooth'' (ordered)
or ``rough'' (massless) phase,
the corresponding zero-temperature spin model must either be
ordered or critical, never disordered.
When the height model is critical,
the long-distance behavior is that of a massless Gaussian
with some ({\em a~priori}\/ unknown) ``stiffness matrix'' ${\bf K} > 0$.
The critical operators can be identified via the height mapping,
and the corresponding critical exponents can be predicted in terms
of ${\bf K}$.
Height representations thus provide a means for recovering
a sort of universality for some (but not all) AF models
and for understanding their critical behavior
in terms of conformal field theory (CFT). 

In particular, on any plane quadrangulation
(i.e., any planar lattice in which all faces are quadrilaterals),
the 3-state Potts AF at $T=0$  
admits a height mapping \cite{Salas_98,Kotecky-Salas-Sokal}.
But is this model critical, or is it ordered?
For the square lattice it is known
\cite{Nijs_82,Kolafa_84,Burton_Henley_97,Salas_98}
that the zero-temperature model is critical, so that $q_c = 3$.
By contrast, for the diced lattice it can be rigorously proven
\cite{Kotecky-Salas-Sokal,Kotecky-Sokal-Swart}
that there is a finite-temperature phase transition,
with an ordered phase at all low temperatures, so that $q_c > 3$
(numerical estimates from transfer matrices yield
$q_c({\rm diced}) \approx 3.45$ \cite{Jacobsen-Salas_unpub}).
Moreover, we have recently \cite{planar_AF_largeq}
given examples of plane quadrangulations
in which $q_c$ takes arbitrarily large values.
It is thus of interest to find conditions on the quadrangulation
telling us whether the 3-state Potts AF at $T=0$ is critical 
or ordered.

In this Rapid Communication we shall propose a criterion,
involving graph duality, that appears 
to give a precise solution to this problem.

Recall first that, for any (finite or infinite) graph $G=(V,E)$
embedded in the plane, the dual graph $G^* = (V^*,E^*)$
is defined by placing a vertex in each face of $G$
and drawing an edge $e^*$ across each edge $e$ of $G$.
Since $G^{**} = G$,
we refer to the pair $(G,G^*)$ as a {\em dual pair}\/.
A graph $G$ is called {\em self-dual}\/ if $G$ is isomorphic to $G^*$.

Now consider a plane quadrangulation $\Gamma = (V,E)$.
Since it is bipartite (say, $V = V_0 \cup V_1$),
we may define sublattices $G_0 = (V_0,E_0)$ and $G_1 = (V_1,E_1)$
by drawing edges across the diagonals of the quadrilateral faces;
it is easy to see that $G_0$ and $G_1$ form a dual pair.
Conversely, given a dual pair $(G_0,G_1)$ of plane graphs,
we can construct a quadrangulation $\scrq(G_0) = \scrq(G_1)$
with the vertex set $V = V_0 \cup V_1$
by connecting each vertex in $G_0$ to the neighboring vertices in $G_1$.
There is thus a one-to-one correspondence between quadrangulations $\Gamma$
and dual pairs of plane graphs $(G_0,G_1)$.
We shall say that the quadrangulation $\scrq(G_0)$ is of 
{\em self-dual type}\/ if $G_0$ is self-dual,
and of {\em  non-self-dual type}\/ otherwise.
For instance, the square lattice is a quadrangulation
of self-dual type (both $G_0$ and $G_1$ are themselves square lattices),
while the diced lattice is a quadrangulation
of non-self-dual type (the sublattices are triangular and hexagonal).

\begin{figure}[t]
\vspace*{0cm} \hspace*{-0cm}
\begin{center}
\includegraphics[width=0.5\columnwidth]{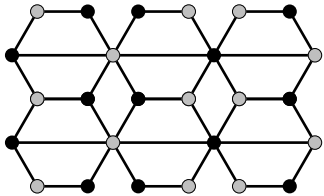}
\end{center}
\vspace*{-5mm}
\caption{
The quadrangulation $\mathcal{Q}(\text{hextri})$. 
}
\label{fig_Qhextri}
\end{figure}

Let us henceforth restrict attention to periodic planar lattices.
It is well known (and obvious) that the square lattice is self-dual;
what seems to be less well known is that there exist
infinitely many examples of self-dual periodic planar lattices
\cite{Ashley_91,Okeeffe_92,Okeeffe_96,Servatius_98,Wierman_06,Scullard_06,%
Ziff_12},
including the ``hextri'' lattice \cite[Figs.~1 and 10]{Okeeffe_92},  
\cite[Fig.~16]{Servatius_98} \cite[Fig.~1b]{Wierman_06},
the ``house'' lattice \cite[Fig.~2]{Okeeffe_92},
the martini-B lattice \cite[Fig.~8]{Scullard_06},
and the cmm-pmm lattice \cite[Fig.~29]{Servatius_98}.
In particular, from each of these lattices we can construct the corresponding
quadrangulation of self-dual type.

In this Rapid Communication we present the results of our study ---
using Monte Carlo (MC), transfer matrices (TM),
and critical polynomials (CP) \cite{JS1} ---
of the 3-state Potts AF on a variety of quadrangulations of both types.
We find empirically, without exception, 
the following behavior:
\begin{conjecture} \label{conj.main}
For the 3-state Potts AF on a (periodic) plane quadrangulation $\Gamma$:

\noindent
(1) If $\Gamma$ is of self-dual type, the model has a zero-temperature 
critical point, so that $q_c = 3$. This critical point has central charge 
$c=1$.

\noindent
(2) If $\Gamma$ is of non-self-dual type, the model has a finite-temperature
phase transition, so that $q_c > 3$. This transition is second-order and 
lies in the universality class of the 3-state Potts ferromagnet.
\end{conjecture}

To our knowledge, prior to this Rapid Communication, only four AF Potts models
on planar lattices with a critical point at $T=0$ were known 
\cite{Henley_unpub,Salas_98}: the square and kagome lattices with $q=3$,
and the triangular lattice with $q=2$ and $q=4$. By contrast,
Conjecture~\ref{conj.main} implies that this phenomenon is not 
so exceptional: There are infinitely many $q=3$ models displaying it!

\begin{figure}[t]
\vspace*{0cm} \hspace*{-0cm}
\begin{center}
\includegraphics[width=0.70\columnwidth]{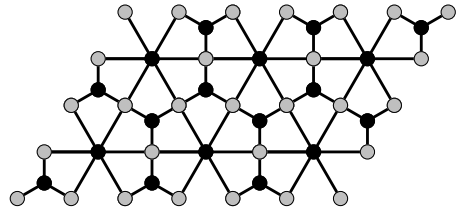} 
\end{center}
\vspace*{-5mm}
\caption{
The quadrangulation $\mathcal{Q}(\text{diced})$, which is also the Laves 
lattice $D(3,4,6,4)$ and is the dual of the ruby lattice.
The black (resp.\/ gray) vertices form a diced (resp.\/ kagome) sublattice.
}
\label{fig_Qdiced}
\end{figure}

We have studied four quadrangulations of self-dual type:
$\mathcal{Q}(\text{hextri})$ [see Fig.~\ref{fig_Qhextri}], 
$\mathcal{Q}(\text{house})$, $\mathcal{Q}(\text{martini-B})$
and $\mathcal{Q}(\text{cmm-pmm})$.
We have also considered four quadrangulations of non-self-dual type:
$\mathcal{Q}(\text{diced})$ (see Fig.~\ref{fig_Qdiced}), 
$\mathcal{Q}(\text{martini})$, $\mathcal{Q}(\text{ruby})$,
and $G''_3$ \cite[Fig.~2(b)]{planar_AF_largeq}. 
As the qualitative behavior of the four lattices within each class turns out
to be the same, we refrain from giving here all the details
%%%%%%%%%%%%%%%%%%%%%%%%%%%%%%%%%%%%%%%%%%%%%%%%%%%%%%%%%%%%%%%%%%%
%(which can be found elsewhere \cite{LDJS_17}),
%%%%%%%%%%%%%%%%%%%%%%%%%%%%%%%%%%%%%%%%%%%%%%%%%%%%%%%%%%%%%%%%%%%
\cite{LDJS_17}, and shall focus on one lattice of each type:
$\mathcal{Q}(\text{hextri})$ and $\mathcal{Q}(\text{diced})$. 

%%\paragraph{Quadrangulations of self-dual type.}
\section{Quadrangulations of self-dual type}

If the quadrangulation is of self-dual type, then we expect the number of
ideal states \cite{Henley_unpub,Salas_98} to be six:
The system must choose which of the two sublattices to order,
and in which of the three possible spin directions.
It is therefore natural to expect (by using universality arguments) that, 
as for the square lattice, there will be a critical point at $T=0$ 
characterized by a CFT with central charge $c=1$ 
\cite{LDJS_17}.

\begin{figure}[t]
\vspace*{0cm} \hspace*{-0cm}
\begin{center}
\includegraphics[width=0.70\columnwidth]{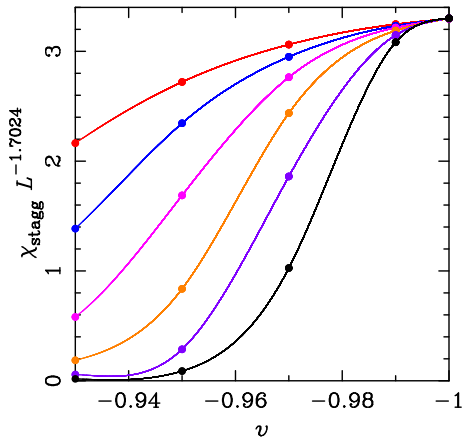} 
\end{center}
\vspace*{-5mm}
\caption{
The scaled staggered susceptibility for the $\mathcal{Q}(\text{hextri})$ 
lattice as a function of $v=e^J-1$ in the region close to $v=-1$.
We show the data (from top to bottom) for $L=32$ (red), $L=64$ (blue),
$L=128$ (pink), $L=256$ (orange), $L=512$ (violet), and $L=1024$ (black).
The curves are spline-interpolation curves to guide the eye.
}
\label{fig_Qhextri_sus}
\end{figure}

We first investigated the 3-state Potts AF by extensive 
MC simulations on lattices of size $L\times L$ unit cells
with periodic boundary conditions (BC),
using the Wang--Swendsen--Koteck\'y (WSK) cluster algorithm \cite{WSK}.
As our lattices are bipartite, this algorithm is known 
to be ergodic even at $T=0$ \cite{Burton_Henley_97,Ferreira_Sokal,Mohar}. 
For each lattice, we measured the staggered and uniform 
susceptibilities, which are expected to diverge at the critical point as 
$\chi_\text{stagg} \sim L^{(\gamma/\nu)_\text{stagg}}$ and
$\chi_\text{u} \sim L^{(\gamma/\nu)_\text{u}}$.  
The qualitative behavior of these susceptibilities is the same for all
four lattices considered here,
but the critical exponents $(\gamma/\nu)_\text{stagg}$ and 
$(\gamma/\nu)_\text{u}$ do depend on the lattice. 
In Fig.~\ref{fig_Qhextri_sus} we show, as an example,
the scaled staggered susceptibility on the $\mathcal{Q}(\text{hextri})$ 
lattice (we use, instead of the standard Potts-model coupling constant~$J$,
the variable $v=e^J-1$).
All the finite-$L$ curves meet at $v=-1$, implying that this point is indeed 
critical. 

The height representation
\cite{Henley_unpub,Kondev_96,Burton_Henley_97,Salas_98} 
relates these susceptibility exponents to the
stiffness $K$ (which is a scalar in this case): 
\begin{equation}
  (\gamma/\nu)_\text{stagg} \;=\; 2 - \frac{\pi}{18K}\,, 
\qquad 
  (\gamma/\nu)_\text{u} \;=\; 2 - \frac{2\pi}{9K}\,.  
\end{equation}
The results for the four lattices studied here, along with the
known exact values for the square lattice \cite{Burton_Henley_97,Salas_98},
are displayed in Table~\ref{tab.gamma_self_dual}.
The value of the stiffness is in all cases much smaller than
the critical value 
$K_c=\pi/2 \approx 1.570\, 796$ where the locking potential becomes marginal,
which separates the rough and smooth phases 
\cite{Henley_unpub,Kondev_96,Salas_98}.

We also found that, for all these lattices,
the WSK algorithm does not suffer from critical slowing-down (CSD).
By measuring the integrated autocorrelation times $\tau_\text{int}$
for the staggered and uniform susceptibilities at $T=0$,
we find that $\tau_{\text{int}} \lesssim 8$ uniformly in $L$. 
This phenomenon also occurs for the square-lattice model
\cite{Salas_98,Ferreira_Sokal}.
We conjecture that the WSK algorithm for the 3-state Potts AF on any
quadrangulation of self-dual type has no CSD.

We also studied the $\mathcal{Q}(\text{hextri})$ lattice by means of
a TM approach. We considered strip graphs of this 
lattice with cylindrical BC and widths $2\le L\le 14$.
In Fig.~\ref{fig_Qhextri}, our TM propagates from left to right.
We measured the free energy (per unit area) $f_L(q)$ at $T=0$ in 
the AF regime for $2 \le q \le 4$. The central charge $c(q)$ can be extracted 
using the standard CFT Ansatz \cite{Bloete_86,Affleck_86} 
\begin{equation}
f_L(q) \;=\; f_\text{bulk}(q) - \frac{ c(q) \, \pi}{6\, L^2} + o(L^{-2}) \,.
\label{cft.ansatz}
\end{equation} 
We first observed that there are parity effects depending on the value of
$L\bmod{4}=0,2$. We then ignored the $o(L^{-2})$ terms, fitted the values 
corresponding to $L,L+4$ to \eqref{cft.ansatz}, and extracted the
estimates $c_L(q)$. This curve exhibits, for each value of $L$, a maximum 
value $c_\text{max}(L)$ at $q=q_\text{max}(L)$. These values are displayed in
Table~\ref{tab.cft.qhextri}, together with our extrapolations to $L=\infty$
(see details of the fits in Ref.~\cite{LDJS_17}). 
These results agree well with our conjecture that the $q=3$ Potts AF 
on $\mathcal{Q}(\text{hextri})$ is critical at $T=0$,
with behavior described by a CFT with $c=1$.

\begin{table}[t]
\centering
\begin{tabular}{llrl}
\hline\hline
\multicolumn{1}{c}{$\Gamma$} &
\multicolumn{1}{c}{$(\gamma/\nu)_\text{stagg}$} &
\multicolumn{1}{c}{$(\gamma/\nu)_\text{u}$} &
\multicolumn{1}{c}{$K$} \\
\hline
$\mathcal{Q}(\text{cmm-pmm})$   & $1.71762(9)$& $0.8691(5)$ & $0.6177(6)$ \\
$\mathcal{Q}(\text{hextri})$    & $1.7024(3)$ & $0.8096(9)$ & $0.5865(6)$ \\
$\mathcal{Q}(\text{house})$     & $1.6978(3)$ & $0.7922(4)$ & $0.5778(8)$ \\
$\mathcal{Q}(\text{martini-B})$ & $1.6882(3)$ & $0.7557(9)$ & $0.5609(6)$ \\
\text{square}                   & \multicolumn{1}{c}{$5/3$}& 
                                  \multicolumn{1}{c}{$2/3$}& 
                                  \multicolumn{1}{c}{$\pi/6$} \\
\hline\hline
\end{tabular}
\caption{Critical exponents $(\gamma/\nu)_\text{stagg}$ and
$(\gamma/\nu)_\text{u}$, and the estimated stiffness~$K$,
for the zero-temperature 3-state Potts AF on the quadrangulations~$\Gamma$
of self-dual type studied in this Rapid Communication.
We include for comparison the exact values
for the square lattice \cite{Salas_98}.
}
\label{tab.gamma_self_dual}
\end{table}

\begin{table}[t]
\centering
\begin{tabular}{lll}
\hline\hline 
$L$ & \multicolumn{1}{c}{$q_\text{max}(L)$} &
      \multicolumn{1}{c}{$c_\text{max}(L)$} \\ 
\hline
2&  3.8544146155 & 0.8508786050 \\
4&  3.2788982545 & 1.0133854086 \\
6&  3.1443621430 & 1.0588380075 \\
8&  3.0975518402 & 1.0554325766 \\
10& 3.0795627986 & 1.0383006482 \\
\hline 
$\infty$ & 3.00(2) & 0.99(2) \\
\hline
\hline
\end{tabular}
\caption{$q_\text{max}(L)$ and $c_\text{max}(L)$ for the
$T=0$ $q$-state Potts AF on the
$\mathcal{Q}(\text{hextri})$ lattice with cylindrical BC
as a function of the width $L$, and the extrapolation to $L=\infty$.
\label{tab.cft.qhextri}
}
\end{table}

%\paragraph{Quadrangulations of non-self-dual type.}
\section{Quadrangulations of non-self-dual type}

When the quadrangulation is of non-self-dual type,
the asymmetry between the two sublattices suggests that at $T=0$
one preferred sublattice will be ordered
(in one of the three possible spin directions)
and the other sublattice disordered (between the other two states).
If this is so, then at $T=0$ there are only three ideal states
each of them with one sublattice ferromagnetically ordered.
Therefore, we have the same ${\mathbb Z}_3$ symmetry
and ground-state degeneracy as for the 
3-state Potts ferromagnet, and hence we expect a finite-temperature 
second-order transition in the universality class of this latter model.
However, a first-order finite-temperature transition is also possible.

In particular, if the two sublattices have unequal vertex densities
(as occurs most often),
then we expect that the sublattice with the smaller (resp.\ larger)
vertex density will be ordered (resp.\ disordered),
as this maximizes the entropy.
The reasoning becomes more subtle, however,
if the two sublattices have equal vertex densities
[as occurs, for instance,
 for $\mathcal{Q}(\text{diced})$ and $\mathcal{Q}(\text{ruby})$]:
Then it is not obvious how the asymmetry alone can drive the phase transition.
For this reason we focus here on $\mathcal{Q}(\text{diced})$. 

\begin{figure}[t]
\vspace*{0cm} \hspace*{-0cm}
\begin{center}
\includegraphics[width=0.70\columnwidth]{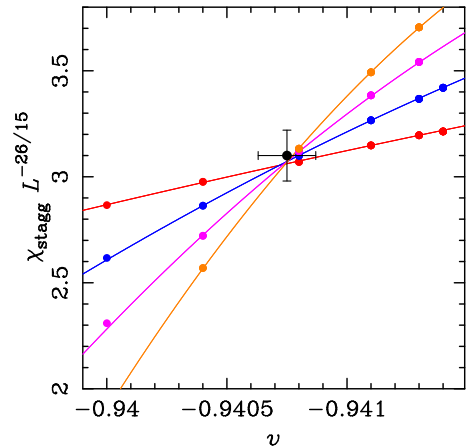} 
\end{center}
\vspace*{-5mm}
\caption{
The scaled staggered susceptibility for the $\mathcal{Q}(\text{diced})$ 
lattice as a function of $v$ in the region close to $v=v_c$.
We show the data for (from top to bottom on the left)
$L=128$ (red), $L=256$ (blue), $L=384$ (pink), and $L=512$ (orange).
The black point shows our best estimate for the parameters,
as well as their corresponding error bars. The curves shown 
are our preferred fits with $(\gamma/\nu)_\text{stagg}$ and $\nu$  
fixed to the exact values $26/15$ and $5/6$.
}
\label{fig_Qdiced_sus}
\end{figure}

Once again we studied the 3-state Potts AF using MC simulations 
on lattices of size $L\times L$ unit cells with periodic BC, 
using the WSK algorithm.
In all cases, we find a finite-temperature critical point.
We followed the practical methods of \cite{Kotecky-Salas-Sokal}
to locate the critical point,
and then fitted our numerical data to the finite-size-scaling (FSS) Ansatz
\begin{multline}
\mathcal{O}_L \;=\; L^{p_\mathcal{O}}\, \bigl[ \mathcal{O}_c + 
   a_1 (v-v_c) L^{1/\nu} + a_2 (v-v_c)^2 L^{2/\nu} \\
                + b_1 L^{-\omega_1} + \cdots \bigr] \:.
\label{FSS.Ansatz}
\end{multline}
For each lattice, we measured the staggered susceptibility $\chi_\text{stagg}$
and the Binder cumulant 
$R_\text{stagg} = \< \bm{\mathcal{M}}_\text{stagg}^4 \>/
                  \< \bm{\mathcal{M}}_\text{stagg}^2 \>^2$.
The qualitative behavior of these observables is the same for
all four lattices considered here.
As an example, we show in Fig.~\ref{fig_Qdiced_sus} the data for the
scaled staggered susceptibility of the $\mathcal{Q}(\text{diced})$ 
lattice, together with our preferred FSS fits
based on the Ansatz~\eqref{FSS.Ansatz} with a varying number of terms.
Since our results for the critical exponents were compatible
with the predicted values $(\gamma/\nu)_\text{stagg} = 26/15$ and $\nu = 5/6$
\cite{Baxter_book},
we then redid the fits fixing these parameters to the predicted values,
in order to obtain improved estimates for $v_c$. Our results are shown in 
Table~\ref{table_results_non_self_dual},
and agree well with the prediction that the model lies
in the universality class of the 3-state Potts ferromagnet.

\begin{table}[t]
\centering
\begin{tabular}{llrrl}
\hline\hline
\multicolumn{1}{c}{$\Gamma$} &
\multicolumn{1}{c}{$v_c$} &
\multicolumn{1}{c}{$(\gamma/\nu)_\text{stagg}$} &
\multicolumn{1}{c}{$\nu$} &
\multicolumn{1}{c}{$R_\text{stagg,c}$} \\
\hline
$\mathcal{Q}(\text{diced})$   & $-0.94075(12)$ & $1.737(6)$             &
                                                 $0.83(7)$  & $1.17(2)$ \\
$\mathcal{Q}(\text{martini})$ & $-0.77454(6)$  & $1.735(4)$             &
                                                 $0.83(3)$  & $1.16(1)$ \\
$\mathcal{Q}(\text{ruby})$    & $-0.95588(9)$  & $1.737(5)$             &
                                                 $0.80(5)$  & $1.15(3)$ \\
$G''_3$                       & $-0.72278(2)$  & $1.736(4)$             &
                                                 $0.82(2)$  & $1.17(1)$ \\
\text{diced}                  & $-0.860599(4)$ & $1.737(4)$             &
                                                 $0.81(2)$  & $1.170(7)$\\
\hline
Prediction                         &                & 
 \multicolumn{1}{c}{$26/15$}  &
 \multicolumn{1}{c}{$5/6$}    &
 \multicolumn{1}{c}{$1.1711(5)$} \\
\hline\hline
\end{tabular}
\caption{Critical temperature $v_c$, critical exponents
$(\gamma/\nu)_\text{stagg}$ and $\nu$, and critical value of the Binder
cumulant $R_\text{stagg,c}$ for the 3-state Potts AF
on quadrangulations~$\Gamma$ of non-self-dual type.
We also include for comparison the results for the diced lattice
$\mathcal{Q}(\text{tri})$ \cite{Kotecky-Salas-Sokal}.
Estimates of $v_c$ are the improved estimates based on fixing
$(\gamma/\nu)_\text{stagg} = 26/15$ and $\nu = 5/6$.
The last line (``Prediction'')
shows the values for the 3-state Potts ferromagnet
\cite{Salas_Sokal_97,Garoni_11}.
}
\label{table_results_non_self_dual}
\end{table}

For the lattice $\mathcal{Q}(\text{diced})$
we have checked directly from the MC simulations that,
even though the two sublattices $G_0$ and $G_1$ have the same vertex density,
it is the diced sublattice $G_0$ (black vertices in Fig.~\ref{fig_Qdiced})
that becomes ordered.
More precisely, the sublattice of $G_0$ consisting of degree-6 vertices
is the one that is most ordered; the two degree-3 sublattices of $G_0$
are slightly more ordered than the three sublattices
of the kagome sublattice $G_1$.
Therefore, although both sublattices $G_0$ and $G_1$
give naively the same entropy density, it is the one having
a sub-sublattice with the largest degree that becomes ordered,
because fluctuations around these three ideal states
maximize the system's entropy density.
A similar phenomenon occurs for the $\mathcal{Q}(\text{ruby})$ lattice
\cite{LDJS_17}.

\begin{table}[t]
\centering
\begin{tabular}{rl}
\hline \hline \\[-4.5mm]
$n$ & \multicolumn{1}{c}{$v_c(n)$} \\
\hline
  2 &  $-0.93449469491145567949$ \\
  4 &  $-0.93889690618313817225$ \\
  6 &  $-0.93976678350525022210$ \\
  8 &  $-0.94017098791714722205$ \\
 10 &  $-0.94038789257375557598$ \\
 12 &  $-0.94051494788357303489$ \\
\hline
$\infty$& $-0.94080(1)$          \\
\hline \hline
\end{tabular}
\caption{Real roots of $P_B(3,v)$, to 20-digit numerical precision, for 
${\cal Q}({\rm diced})$. We show the unique real root 
$v_c(n)$ in the AF interval $v\in [-1,0)$, for $n \times \infty$ bases,
together with the extrapolation to $n=\infty$  
(where we used
exponents in the range $1.2$--$1.5$, which are much smaller than those for
the ferromagnetic models investigated in Refs.~\cite{JS1,JS2,JS3}).}
\label{tab.PB_Q_diced}
\end{table}

On all these lattices (as well as on the diced lattice 
\cite{Kotecky-Salas-Sokal}), the WSK algorithm suffers from CSD,
with dynamic critical exponents
$z_{\textrm{int},\scrm_\textrm{stagg}^2} = 0.50(1)$ and 
$z_{\textrm{int},\scrm_\textrm{u}^2} = 0.48(1)$. 
If these exponents are in fact equal, then our preferred estimate
(taking into account the statistical non-independence of the two
estimates) would be $z_{\rm int} = 0.49(2)$.  This is compatible
with the exponent $z_{\textrm{int},\scrm^2} = 0.475(6)$
found in the Swendsen--Wang (SW) algorithm \cite{Swendsen-Wang}
for the 3-state Potts ferromagnet \cite{Salas_Sokal_97,Garoni_11}.

Finally, we have applied the CP method \cite{JS1,JS2,JS3} to study the 
location of the critical point for the 
3-state Potts AF on the $\mathcal{Q}(\text{diced})$ lattice. 
We computed the CP $P_B(3,v)$ for some bases $B$ 
that admit a four-terminal representation (to be able to use the TM method 
of Ref.~\cite{JS2}). In particular, to compute the estimates of $v_c$ 
shown in Table~\ref{tab.PB_Q_diced},
we have used the more powerful eigenvalue method of Ref.~\cite{JS3}, 
which allows 
us to use bases of size $n\times m$ in the limit $m\to\infty$. 
The last row of Table~\ref{tab.PB_Q_diced} shows the extrapolation to
$n=\infty$ using Monroe's implementation of the Bulirsch--Stoer \cite{BS}
extrapolation scheme. This result agrees within errors with the MC estimate,
but it is more precise. 

\section{Conclusions}

We have studied the 3-state Potts AF on four quadrangulations of self-dual
type, and on four quadrangulations of non-self-dual type
(including two with equal vertex densities on the two sublattices),
by extensive computations using MC simulations, TM computations,
and the CP method. In all cases, we have found a perfect agreement with 
Conjecture~\ref{conj.main}. Our findings provide very strong empirical
support for the validity of this criterion.
However, we do not want to exclude the possibility
that for some lattices of non-self-dual type the finite-temperature
transition might be first-order.

As a side result, we have also found that the WSK algorithm
has no CSD when simulating the 3-state Potts AF
on any quadrangulation of self-dual type, while it has CSD 
(compatible with the dynamic universality class of the SW algorithm
for the 3-state ferromagnet)
on any quadrangulation of non-self-dual type.

\bigskip

\begin{acknowledgments}
We thank Kun Chen and Yuan Huang for their participation in an early stage 
of this work.
This work was supported in part by 
%YD 
the Natural Science Foundation of China Grants No.~11625522, 
%J-PL
No.~11405003  and No.~11774002, 
the Key Projects of Anhui Province University Outstanding Youth Talent 
Support Program Grant No.~gxyqZD2017009, 
%YD 
the Ministry of Science and Technology of China Grant No.~2016YFA0301600,  
% JLJ
the Institut Universitaire de France, 
the European Research Council through the Advanced Grant NuQFT, 
% JS, JLJ, ADS
the Spanish MINECO FIS2014-57387-C3-3-P and 
MINECO/AEI/FEDER, UE  FIS2017-84440-C2-2-P grants, and 
% ADS 
EPSRC Grant No.~EP/N025636/1. 
\end{acknowledgments}

%%%%%%%%%%%%%%%%%%%%%%%%%%%%%%%%%%%%%%%%%%%%%%%%%%%%%%%%%%%%%%%%%%%%%%


\begin{thebibliography}{99}

\bibitem{Kramers_41}  H.A. Kramers and G.H. Wannier, Phys. Rev. {\bf 60}, 252
   (1941).

\bibitem{Savit_80}  R. Savit, Rev. Mod. Phys. {\bf 52}, 453 (1980).

\bibitem{Polchinski_96}  J. Polchinski, Rev. Mod. Phys. {\bf 68}, 1245 (1996).

\bibitem{Potts_52}  R.B. Potts,
   Proc. Cambridge Philos. Soc. {\bf 48}, 106 (1952).

\bibitem{Wu_82+84}  F.Y. Wu, Rev. Mod. Phys. {\bf 54}, 235 (1982);
   {\bf 55}, 315 (E) (1983); F.Y. Wu, J. Appl. Phys. {\bf 55}, 2421 (1984).

\bibitem{Baxter_book} R.J. Baxter, {\em Exactly Solved Models in Statistical
   Mechanics}\/ (Academic Press, London--New York, 1982).

\bibitem{Nienhuis_84}  B. Nienhuis, J. Stat. Phys. {\bf 34}, 731 (1984).

\bibitem{DiFrancesco_97}  P. Di Francesco, P. Mathieu and D. S\'en\'echal,
   {\em Conformal Field Theory}\/ (Springer-Verlag, New York, 1997).

\bibitem{Henley_unpub}  C.L. Henley, Discrete Spin Models with ``Height''
   Representations and Critical Ground States, unpublished manuscript 
   (September, 1993).

\bibitem{Kondev_96} J. Kondev and C.L. Henley, Nucl. Phys. B {\bf 464}, 540
   (1996), \arxiv{cond-mat/9511102}. 

\bibitem{Burton_Henley_97}  J.K. Burton Jr. and C.L. Henley,
   J. Phys. A {\bf 30}, 8385 (1997), \arxiv{cond-mat/9708171}.

\bibitem{Salas_98}  J. Salas and A.D. Sokal, J. Stat. Phys {\bf 92},
   729 (1998), \arxiv{cond-mat/9801079} and references therein. 

\bibitem{Jacobsen_09}  J.L. Jacobsen, in {\em Polygons, Polyominoes and 
   Polycubes}\/, edited by A.J. Guttmann, Lecture Notes in Physics, Vol.~775
   (Springer, Dordrecht, 2009), Chapter~14.

\bibitem{Kotecky-Salas-Sokal}  R. Koteck\'y, J. Salas, and A.D. Sokal,
   Phys. Rev. Lett. {\bf 101}, 030601 (2008), \arxiv{0802.2270}.

\bibitem{Nijs_82}  M.P.M. den Nijs, M.P. Nightingale, and M. Schick,
   Phys. Rev. B {\bf 26}, 2490 (1982).

\bibitem{Kolafa_84} J. Kolafa, J. Phys. A {\bf 17}, L777 (1984).

\bibitem{Kotecky-Sokal-Swart}  R. Koteck\'y, A.D. Sokal, and J.M. Swart,
   Commun. Math. Phys. {\bf 330}, 1339 (2014), \arxiv{1205.4472}.

\bibitem{Jacobsen-Salas_unpub}  J.L. Jacobsen and J. Salas, unpublished (2008).

\bibitem{planar_AF_largeq}  Y. Huang, K. Chen, Y. Deng, J.L.Jacobsen,
   R. Koteck\'y, J. Salas, A.D. Sokal, and J.M. Swart,
   Phys. Rev. E {\bf 87}, 012136 (2013), \arxiv{1210.6248}.

\bibitem{Ashley_91}  J. Ashley, B. Gr\"unbaum, G.C. Shephard, and 
   W. Stromquist,
   %% Self-duality groups and ranks of self-dualities,
   in {\em Applied Geometry and Discrete Mathematics}\/
   %% DIMACS Series in Discrete Mathematics and Theoretical Computer Science,
   %% vol.~4
   (American Mathematical Society, Providence RI, 1991), pp.~11--50.

\bibitem{Okeeffe_92}  M. O'Keeffe, Aust. J. Chem. {\bf 45}, 1489 (1992).

\bibitem{Okeeffe_96}  M. O'Keeffe and B.G. Hyde,
   {\em Crystal Structures I.~Patterns and Symmetry}\/
   (Mineralogical Society of America, Washington DC, 1996),
   Section~5.3.7. Available on-line at
   \url{http://www.public.asu.edu/~rosebudx/okeeffe.htm}

\bibitem{Servatius_98}  B. Servatius and H. Servatius,
   %% Symmetry, automorphisms, and self-duality of infinite planar graphs
   %% and tilings,
   in {\em Proceedings of the International Scientific Conference on
    Mathematics}\/ (\v{Z}ilina, 30 June -- 3 July 1998),
    edited by V.~B\'alint
    (University of \v{Z}ilina, \v{Z}ilina, 1998), pp.~83--116.
   Available on-line at
   \url{http://users.wpi.edu/~bservat/self5.html}

\bibitem{Wierman_06}  J.C.~Wierman,
   %% Construction of infinite self-dual graphs,
   in {\em Proceedings of the 5th Hawaii International Conference on
   Statistics, Mathematics and Related Fields}\/ (2006) [CD-ROM]. 
   %%Available on-line at
   %%\url{http://www.ams.jhu.edu/~wierman/publications.htm}.

\bibitem{Scullard_06}  C.R.~Scullard, Phys. Rev. E {\bf 73}, 016107 (2006),
   \arxiv{cond-mat/0507392}.

\bibitem{Ziff_12}  R.M. Ziff, C.R. Scullard, J.C. Wierman, and M.R.A. Sedlock,
   J. Phys. A {\bf 45}, 494005 (2012), \arxiv{1210.6609}.

\bibitem{JS1}  J.L. Jacobsen and C.R. Scullard, J. Phys. A {\bf 45}, 494003 
   (2012), \arxiv{1204.0622};
   C.R. Scullard and J.L. Jacobsen, J. Phys. A {\bf 45}, 494004 (2012), 
   \arxiv{1209.1451};
   J.L. Jacobsen and C.R. Scullard, J. Phys. A {\bf 46}, 075001 (2013), 
   \arxiv{1211.4335};    
   C.R. Scullard and J.L. Jacobsen, J. Phys. A {\bf 49}, 125003 (2016),
   \arxiv{1511.04374}.

\bibitem{LDJS_17} J.P. Lv, Y. Deng, J.L. Jacobsen, and J. Salas, 
   The three-state Potts antiferromagnet on plane quadrangulations,
   \arxiv{1804.08911}. 

\bibitem{WSK} J.-S. Wang, R.H. Swendsen, and R. Koteck\'y,
   Phys. Rev. Lett. {\bf 63}, 109 (1989); Phys. Rev. B {\bf 42}, 2465 (1990).

\bibitem{Ferreira_Sokal} S.J. Ferreira and A.D. Sokal,
   J. Stat. Phys. {\bf 96}, 461 (1999), \arxiv{cond-mat/9811345}.

\bibitem{Mohar} B. Mohar,
  %% Kempe equivalence of colorings,
  in {\em Graph Theory in Paris}\/,
  edited by J.A.~Bondy {\em et al.}\/
  %% J. Fonlupt, J.L. Fouquet, J.-C. Fournier and J. Ram\'{\i}rez Alfons\'{\i}n
  (Birkh\"auser, Basel, 2007), pp.~287--297.

\bibitem{Bloete_86} H.W.J. Bl\"ote, J.L. Cardy, and M.P. Nightingale,
  Phys. Rev. Lett. {\bf 56}, 742 (1986).

\bibitem{Affleck_86} I. Affleck,
  Phys. Rev. Lett. {\bf 56}, 746 (1986).

\bibitem{Swendsen-Wang} R.H. Swendsen and J.-S. Wang,
   Phys. Rev. Lett. {\bf 58}, 86 (1987).

\bibitem{Salas_Sokal_97} J. Salas and A.D. Sokal, J. Stat. Phys. {\bf 87},
     1 (1997), \arxiv{hep-lat/9605018}.

\bibitem{Garoni_11} T.M. Garoni, G. Ossola, M. Polin, and A.D. Sokal,
     J.~Stat. Phys. {\bf 144}, 459 (2011), \arxiv{1105.0373}.

\bibitem{JS2} J.L. Jacobsen, J. Phys. A {\bf 47}, 135001 (2014), 
     \arxiv{1401.7847}. 

\bibitem{JS3} J.L. Jacobsen, J. Phys. A {\bf 48}, 454003 (2015),
     \arxiv{1507.03027}.

\bibitem{BS} R. Burlisch and J. Stoer, Numer. Math. {\bf 6}, 413 (1964);
     J.L. Monroe, Phys. Rev. E {\bf 65}, 066116 (2002).

\end{thebibliography}
\end{document}